\documentclass[aps,pra,reprint,amsmath,amssymb,superscriptaddress,showpacs,nobibnotes]{revtex4-1}
\usepackage{graphicx,SIunits}
\usepackage{bm}     
\usepackage{hyperref} 
\usepackage{dsfont}    
\usepackage{color}
\usepackage[normalem]{ulem}
\DeclareMathOperator{\Tr}{Tr}

\begin{document}


\title{Revealing hidden quantum steerability using local filtering operations}



\author{Tanumoy Pramanik}
\email{tanu.pra99@gmail.com}
\affiliation{Center for Quantum Information, Korea Institute of Science and Technology (KIST), Seoul, 02792, Republic of Korea}


\author{Young-Wook Cho}
\affiliation{Center for Quantum Information, Korea Institute of Science and Technology (KIST), Seoul, 02792, Republic of Korea}

\author{Sang-Wook Han}
\affiliation{Center for Quantum Information, Korea Institute of Science and Technology (KIST), Seoul, 02792, Republic of Korea}

\author{Sang-Yun Lee}
\affiliation{Center for Quantum Information, Korea Institute of Science and Technology (KIST), Seoul, 02792, Republic of Korea}

\author{Yong-Su Kim}
\email{yong-su.kim@kist.re.kr}
\affiliation{Center for Quantum Information, Korea Institute of Science and Technology (KIST), Seoul, 02792, Republic of Korea}
\affiliation{Division of Nano \& Information Technology, KIST School, Korea University of Science and Technology, Seoul 02792, Republic of Korea}

\author{Sung Moon}
\affiliation{Center for Quantum Information, Korea Institute of Science and Technology (KIST), Seoul, 02792, Republic of Korea}
\affiliation{Division of Nano \& Information Technology, KIST School, Korea University of Science and Technology, Seoul 02792, Republic of Korea}

\date{\today} 

\begin{abstract}
\noindent Nonlocal quantum correlation is at the heart of bizarre nature of quantum physics. While there are various classes of nonlocal quantum correlation, steerability of a quantum state by local measurements provides unique operational features. Here, we theoretically and experimentally investigate the `hidden' property of quantum steerability. In particular, we find that there are initially unsteerable states which can reveal the steerability by using local filters on individual quantum systems. It is remarkable that a certain set of local filters are more effective on revealing steerability than Bell nonlocality whereas there exists another set of filters that is more effective on revealing Bell nonlocality than steerability. This result suggests the structural difference between entanglement and steerability. Finally, we present a counter-intuitive result that mixed states originating from non-maximally pure entangled states can have hidden steerability while the mixed state from a maximally pure entangled state fails to show steerability.
\end{abstract}

\keywords{Hidden nonlocality, Hidden steerability, Steering, Quantum key distribution, }

\maketitle
\section{Introduction}

Since Einstein, Podolsky and Rosen suggested the EPR paradox, nonlocal quantum correlation has been understood as one of the most peculiar nature of quantum physics~\cite{EPR}. While entanglement has attracted a lot of attention, it has been known that there are other classes of nonlocal quantum correlation such as Bell nonlocality and quantum steerability~\cite{Bell, CHSH, Jones07_1,Jones07_2, Rev_BN}. Among them, steerability of quantum states by local measurements, which was first suggested by Schr\"{o}dinger~\cite{Schrodinger_1}, provides unique operational features of quantum theory. 

Although the concept of quantum steerability has been recognized from the early age of quantum physics, its definition and mathematical description have been relatively recently introduced~\cite{Jones07_1,Jones07_2}. The operational meaning of quantum steerability can be understood as follows.
{\color{black} Let us assume that Alice prepares two systems $A$ and $B$ in an entangled state and sends the system $B$ to Bob. Here, Bob does not trust Alice and only agrees that he receives quantum system $B$. Bob will be convinced by Alice that the prepared state is entangled when Alice can control $B$ in arbitrary quantum states, e.g., Alice can collapse the state of  $B$ in the eigenstates of two non-commuting observables with the precision larger than that allowed by uncertainty relation. In this case, Bob system $B$ {\color{black} cannot be} described by local hidden state (LHS) model.}

This interpretation leads a remarkable consequence that verification of quantum steerability can be considered as verification of entanglement with an untrusted party (here, Alice), and thus, all the steerable states are entangled. However, it is notable that not all the entangled states guarantee steerability~\cite{Jones07_1,Jones07_2}. On the other hand, Bell nonlocality, another nonlocal quantum correlation determined by local hidden variable (LHV) model, guarantees the steerability, while the reverse is not true~\cite{Saunders}. Therefore, quantum steerability lies between entanglement and Bell nonlocality.


{\color{black} Besides the entanglement verification, the monogamy relation of quantum steering~\cite{PE_St1} provides the security of quantum key distribution when one of the {\color{black} systems} (here, system A) is not trusted as quantum systems.} This scenario is known as one-sided device independent quantum key distribution (1s DI-QKD)~\cite{1sDIQKD}. Similarly, Bell nonlocality guarantees the security of quantum key distribution in fully device independent manner known as DI-QKD~\cite{DIQKD_1,DIQKD_2,DIQKD_3,DIQKD_4}. Note that the implementation of 1s DI-QKD is more practical than that of DI-QKD since quantum steerability is more robust against environmental noise and particle loss than Bell nonlocality~\cite{bennet12,wittmann12,smith12}. 

The characterization and quantification of quantum steerability are, therefore, of great importance not only for fundamental quantum information science but also for applications in quantum communication. Due to difficulty in finding LHS model for mixed entangled states, however, characterizing steerability is challenging~\cite{PE_St3, CHSH-Steer}. It becomes even more complicated since there are unsteerable states that become steerable when local measurements are performed on multiple copies of the pairs rather than a single copy~\cite{Steer_Acti,quintino16}. Note that this phenomenon is known as super-activation of quantum steerability, and a similar phenomenon can also be found in Bell nonlocality~\cite{Bell_Acti_1,Bell_Acti_2,Bell_Acti_3}. 

In this paper, motivated by hidden Bell nonlocality~\cite{HN_Bell1,HN_Bell2,kwiat01,verstraete02,HN_Bell3,HN_Bell4,Hirsch13},  we theoretically and experimentally investigate the revealing of `hidden' quantum steerability. Note that a recent theoretical study proposes that hidden steerability can be found by reducing the dimension of Hilbert space~\cite{quintino15}. In the present work, we reveal hidden steerability with the help of local filters associated with particle loss while the dimension of Hilbert space is unchanged. In particular, we present that there exist bipartite states, which are initially {\it not} steerable (i.e., are explained by LHS model), can become steerable with the help of local filtering operations on the individual systems. Remarkably, we found that there exists a certain set of local filters that is more effective on {\color{black} revealing steerability than Bell nonlocality and vice versa}.
We also present a counter-intuitive result that mixed states which are obtained from non-maximally pure entangled states can show hidden steerability for wider range of a mixing parameter than the mixed state originated from a maximally pure entangled state.




\section{Theory}

In order to investigate different hidden quantum correlations, let us assume that Alice and Bob share a bipartite state of
\begin{equation}
\rho = p\, |\psi\rangle_{\theta}\langle\psi| + (1-p)\, \rho_{A} \otimes \frac{I}{2},
\label{State}
\end{equation}
where $|\psi\rangle_{\theta}=\cos\theta\,|00\rangle+\sin\theta\,|11\rangle$, $\rho_A=\Tr_B\left[|\psi\rangle_{\theta}\langle\psi|\right]$, and $I$ denotes the identity $2\times2$ matrix. Note that the parameters $p$ and $\theta$ hold the following conditions of $p\in[0,1]$, and $\theta\in[0,\frac{\pi}{4}]$, respectively. For simplicity, we define a parameter $\gamma$ to determine the ratio between $|00\rangle$ and $|11\rangle$ as 
$\gamma=\frac{\cos^2\theta}{\sin^2\theta}=\cot^2\theta$.

{\color{black}The  state $\rho$ can be prepared as follows.} Alice prepares a bipartite entangled state of $|\psi\rangle_{\theta}$, and sends one of the particles to Bob while keeping the other. If the transmission channel has depolarizing noise with probability of $1-p$, the shared two-qubit state will become $\rho$ as in Eq.~(\ref{State}). 

Let us consider entanglement, Bell nonlocality and steerability of the state $\rho$. In this work, we restrict our study of Bell nonlocality to Bell-CHSH inequality with two measurement settings and two measurement outcomes for each party~\cite{BE3322}. {\color{black} In the following, we only consider the steerability from Alice to Bob}. Here, we only provide the results of the estimation with brief explanation. Detailed estimation procedure can be found in the Supplemental materials~\cite{Supplement}.  Note that Bell nonlocality can be enhanced by increasing the number of measurement settings~\cite{BNnm1,BNnm2}, however, it is often impractical. For instance, by increasing the number of measurement settings from 2 to 465, the lower bound of Bell nonlocality of the state Eq.~(\ref{State}) with $\theta=\pi/4$ changes from 0.707 to 0.706~\cite{BNnm1}.

Considering concurrence of the state $\rho$~\cite{Concurrence_1,Concurrence_2}, {\color{black}it} is entangled for $p>1/3$~\cite{Supplement}. According to the Horodecki criterion~\cite{Horo_Cri,Horo_Cri_2}, the state $\rho$ is Bell nonlocal if $p > \frac{1}{\sqrt{1+\sin^22\theta}}$~\cite{Supplement}. Note that, for the scenario of two measurement settings on each $2\times 2$ and $2\times 3$ dimensional systems, the Horodecki criterion is necessary and sufficient for testing Bell nonlocality.

Unlike entanglement and Bell nonlocality, it is not straightforward to capture all the possible steerability of a given quantum state since it is difficult to discriminate whether it can be demonstrated by LHS model or not~\cite{Jones07_1}. However, the sufficient criterion for unsteerability from Alice to Bob of the state $\rho$ has been derived in the Ref.~\cite{PE_St3} as  
\begin{eqnarray}
T_U=\frac{A^2+2 |B|}{2} \leq \frac{1}{2},
\label{Steer_Bowel}
\end{eqnarray}
where $A=\frac{(1-p^2) \cos2\theta}{1-p^2\cos^22\theta}$, and $B=\frac{p(1-\cos^22\theta)}{1-p^2\cos^22\theta}$. In other words, Eq.~(\ref{Steer_Bowel}) guarantees unsteerability of $\rho$ from Alice to Bob. 
Note, however, that Eq.~(\ref{Steer_Bowel}) cannot be directly tested with experiment. In the experiment, we have verified unsteerability with the experimentally reconstructed density matrices via quantum state tomography.

For experimental investigation of steerability, we employ the fine-grained steering criterion which is presented in Method in detail~\cite{PE_St1}. Note that, according to the criterion, {\color{black} Alice's steerability on Bob's state} can be experimentally verified by observing the violation of the steering inequality of
\begin{equation}
T=\frac{1}{2} \left[P(b_{\mathcal{B}_1}|a_{\mathcal{A}_1}) + P(b_{\mathcal{B}_2}|a_{\mathcal{A}_2}) \right] \leq \frac{3}{4},
\label{FUR_St}
\end{equation}
where $P(b_{\mathcal{B}}|a_{\mathcal{A}})$ denotes the conditional probability and $\mathcal{A}_{1,2}$ and $\mathcal{B}_{1,2}$ are non-commuting measurement bases chosen by Alice and Bob, respectively. Here, for simplicity, we choose $\mathcal{A}_1,\mathcal{B}_1=\sigma_z$ and $\mathcal{A}_2,\mathcal{B}_2=\sigma_x$, respectively. Note that the above measurement bases are {\color{black} optimized, i.e., they are sufficient to reveal steerability of pure entangled states $|\psi\rangle_\theta$ for all values of $\theta$}~\cite{PE_St1}. It is difficult to find if they are  also optimal for the state  $\rho$, however, we will show that they are sufficiently good to examine the steerability of $\rho$ by comparing with the criterion of Eq.~(\ref{Steer_Bowel}). With these measurement bases, the steering parameter $T$ which is given in the left-hand side of Eq.~(\ref{FUR_St}) becomes $T=(2\,+\,p+\,p\sin2\theta)/4$ for the state $\rho$. Hence, the state $\rho$ is steerable if
\begin{equation}
p>\frac{1}{1\,+\,\sin2\theta}.
\label{Steer_Sigma}
\end{equation}


In order to reveal different hidden quantum correlations, we first examine the local filters of Alice and Bob given as
\begin{eqnarray}
\mathcal{F}_A = \begin{pmatrix}
\frac{1}{\cos\theta} & 0\\
0 & \frac{1}{\sin\theta}
\end{pmatrix},~
\mathcal{F}_B = \begin{pmatrix}
1 & 0\\
0 & 1
\end{pmatrix}.
\label{Filter}
\end{eqnarray}
After the local filters, the initial state $\rho$ becomes a Werner state of $\rho_{\mathcal{F}}=\rho_{\pi/4}= p \,|\psi\rangle_{\frac{\pi}{4}}\langle\psi | + (1-p) \,\frac{I}{2}\otimes \frac{I}{2}$. {\color{black} The state $\rho_{\mathcal{F}}$ is entangled for $p>1/3$ and  Bell nonlocal for $p>1/\sqrt{2}$~\cite{Supplement}. Therefore, the state $\rho$ shows hidden Bell nonlocality with local filters of $\mathcal{F}_{A,B}$ for $\frac{1}{\sqrt{2}} <p \leq \frac{1}{\sqrt{1+\sin^22\theta}}$. Note that the above choice  of $\mathcal{F}_{A,B}$ are optimal to reveal hidden Bell nonlocality~\cite{OBF}}.

Since the final state $\rho_{\mathcal{F}}$ is a Werner state, it is steerable for $p>1/2$~\cite{Jones07_1,Jones07_2,PE_St1}. Therefore, local filters of Eq.~(\ref{Filter}) reveal hidden steerability for $1/2< p \leq \frac{1}{1+\sin2\theta}$ of the state $\rho$. Note that for the region of $\frac{1}{2}<p<\frac{1}{\sqrt{2}}$, $\rho$ shows hidden steerability but not Bell nonlocality. This result clearly shows that hidden steerability is distinct from hidden Bell nonlocality. 

Now, let us consider another local filters of
\begin{eqnarray}
\mathcal{G}_A = \begin{pmatrix}
\frac{1}{\sqrt{\cos\theta}} & 0\\
0 & \frac{1}{\sqrt{\sin\theta}}
\end{pmatrix}, ~~~ \mathcal{G}_B = \begin{pmatrix}
\frac{1}{\sqrt{\sin\theta}} & 0\\
0 & \frac{1}{\sqrt{\cos\theta}}
\end{pmatrix}.\hspace{0.2cm}
\label{Filter_St}
\end{eqnarray}
The state $\rho_{\mathcal{G}}$ is entangled for $p>1/3$ and  Bell nonlocal for  $\left(p^2-1\right) \csc \theta  \sec \theta +5 p^2+(p+1)^2 \csc ^2 2 \theta -2 p+1>\frac{1}{2}[(p+1) \csc \theta \sec \theta -2 p+2]^2$~\cite{Supplement}.
Thus, the state $\rho_{\mathcal{G}}$ is steerable for $p>\delta$, where $\delta=\Big[\cot \theta-\cos^2\theta (\cot \theta+2)-7 \sin \theta \cos\theta+\sqrt{ (6\sin 2 \theta +5 \sin4\theta+6 \cos 2 \theta+10)}\Big]\Big/\Big[2 (-2  \sin2\theta +\cos2 \theta+2)\Big]$~\cite{Supplement}. 
Therefore, the state $\rho$ shows hidden steerability for $\delta < p \leq \frac{1}{1+\sin2\theta}$. Note that it is hard to find that if $\mathcal{G}_{A,B}$ {\color{black} are} optimal for revealing hidden steerability.

\begin{figure}[b!]
\includegraphics[width=3.4in]{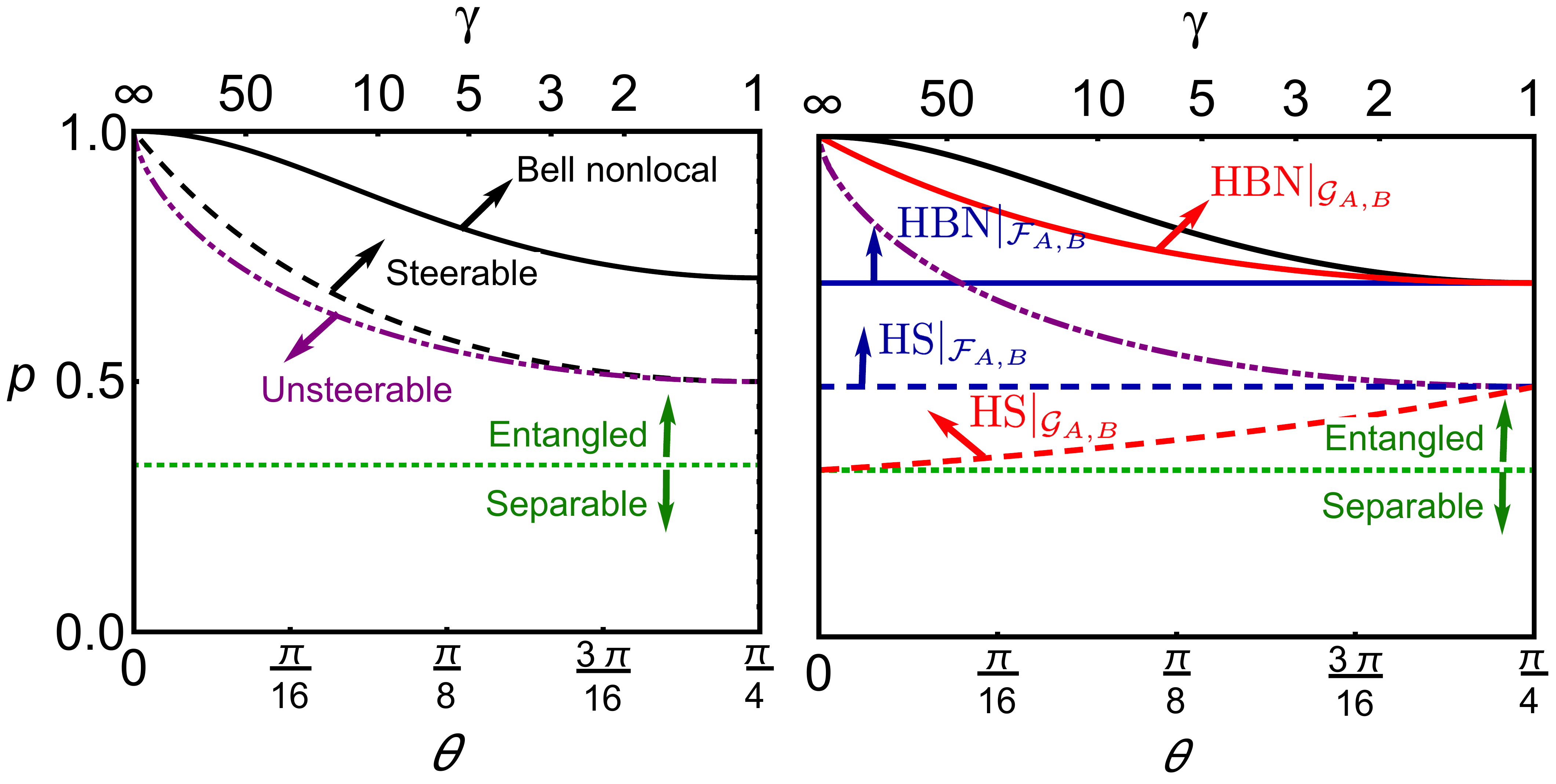}
\caption{ (a) The regions of various quantum correlations in terms of state parameters $\{p,\,\theta\}$ (or equivalently $\{p,\,\gamma\}$ where $\gamma$ is defined in the paragraph {\color{black} following}  the Eq.~(\ref{State})) for the bipartite state $\rho$ of Eq.~(\ref{State}). The boundaries of Bell nonlocality and quantum steerability are presented by solid and dashed black lines, respectively. The purple line denotes the unsteerability boundary determined by Eq.~(\ref{Steer_Bowel}). (b) The regions of hidden quantum correlations revealed by different local filters. ${\rm HBN}|_{\mathcal{H}_{A,B}}$ and ${\rm HS}|_{\mathcal{H}_{A,B}}$ denote hidden Bell nonlocal and hidden steerable with local filters of $\mathcal{H}_{A,B}\in\{\mathcal{F}_{A,B},\mathcal{G}_{A,B}\}$, respectively. In order to clearly show hidden steerability, we present the unsteerability boundary of the original state $\rho$ given by Eq.~(\ref{Steer_Bowel}) with a purple line.}
\label{Fig_1}
\end{figure}


We summarize different regions of various quantum correlations of the state $\rho$ of Eq.~(\ref{State}) with respect to the parameters $\theta$ and $p$ in Fig.~\ref{Fig_1}. First, let us first consider different quantum correlations of the state $\rho$, see Fig.~\ref{Fig_1}(a). It is remarkable that the steering boundaries of Eqs.~(\ref{Steer_Bowel}) and (\ref{Steer_Sigma}) are comparable. It means that the set of measurement bases of $\{\sigma_x,\sigma_z\}$ for the steering inequality test provides a relatively close steering boundary to its optimum. It is interesting that for a Bell state, i.e., $\gamma=1$, the boundaries of Eq.~(\ref{Steer_Bowel}) and (\ref{Steer_Sigma}) are identical and both criteria become necessary and sufficient. However, the fine-grained steering criterion of Eq.~(\ref{Steer_Sigma}) fails to detect steerability of $\rho$ having $\{p,\theta\}$ lies between the two above boundaries.


Now, let us turn our interest to the hidden quantum correlations revealed by local filtering operations, see Fig.~\ref{Fig_1}(b). In order to clearly show the hidden steerability, we present the sufficient criterion of unsteerability of Eq.~(\ref{Steer_Bowel}) with a purple dot-dashed line. If the steering boundaries of the post-filtered states $\rho_{\mathcal{F}}$ and $\rho_{\mathcal{G}}$ are found below this sufficient criterion, one can be sure that the local filtering operations successfully reveal hidden steerability. As presented by dashed blue and red lines, it is clear that both $\mathcal{F}_{A,B}$ and $\mathcal{G}_{A,B}$ can reveal hidden steerability.

As shown in the paragraph following Eq.~(\ref{Filter}), applying $\mathcal{F}_{A,B}$ makes the initial state $\rho$ which has biased $\gamma>1$ to a Werner state $\rho_{\pi/4}$ or equivalently $\gamma=1$ while keeping the parameter $p$ unchanged. Therefore, $\mathcal{F}_{A,B}$ uncover hidden Bell nonlocality and steerability of $\rho$ up to those of the Werner states $\rho_{\pi/4}$. In Fig.~\ref{Fig_1}(b), these are represented as horizontal blue lines.

The results of applying $\mathcal{G}_{A,B}$ are more interesting. Comparing to $\mathcal{F}_{A,B}$, the area for hidden Bell nonlocality is reduced while that of hidden steerability is enlarged. Therefore, if we determine the effectiveness of local filters as the area of uncovered nonlocal correlation region, $\mathcal{G}_{A,B}$ are less effective on revealing Bell nonlocality than $\mathcal{F}_{A,B}$. However, $\mathcal{G}_{A,B}$ are more effective on uncovering steerability than $\mathcal{F}_{A,B}$. These results intimate the structural difference between Bell nonlocality and steering.

It is remarkable that as $\theta\rightarrow0$ (equivalently, $\gamma\rightarrow\infty$), the region of hidden steerability revealed by $\mathcal{G}_{A,B}$ is broadened while the regions of other quantum correlation contracted or remain the same. Note that the amount of entanglement decreases as $\theta\rightarrow0$ for a fixed $p$. Therefore, it suggests a counter-intuitive result that mixed states from non-maximally entangled states can have hidden steerability for wider range of the mixing noisy parameter $p$ than the mixed state originated from a maximally entangled state.

\section{Experiment}


For experimental verification of our findings, we have conducted experiments with polarization entangled photon pairs from spontaneous parametric down conversion. The Bell nonlocality and steerability are tested with the CHSH and fine-grained steering inequalities. The sufficient condition of unsteerability of Eq.~(\ref{Steer_Bowel}) is also tested experimentally via quantum state~tomography~\cite{qst1,qst2}. {\color{black} Note that the Bell parameter $S$ ({\color{black} for definition, see the supplemental material~\cite{Supplement}}), $T$ and $T_U$ are not directly measurable and they are calculated from the probability distribution of the measurement outcomes of different observables on the respective systems.} The details of experiment can be found in the supplemental material~\cite{Supplement}.

\begin{figure}[t]
\includegraphics[width=3.3in]{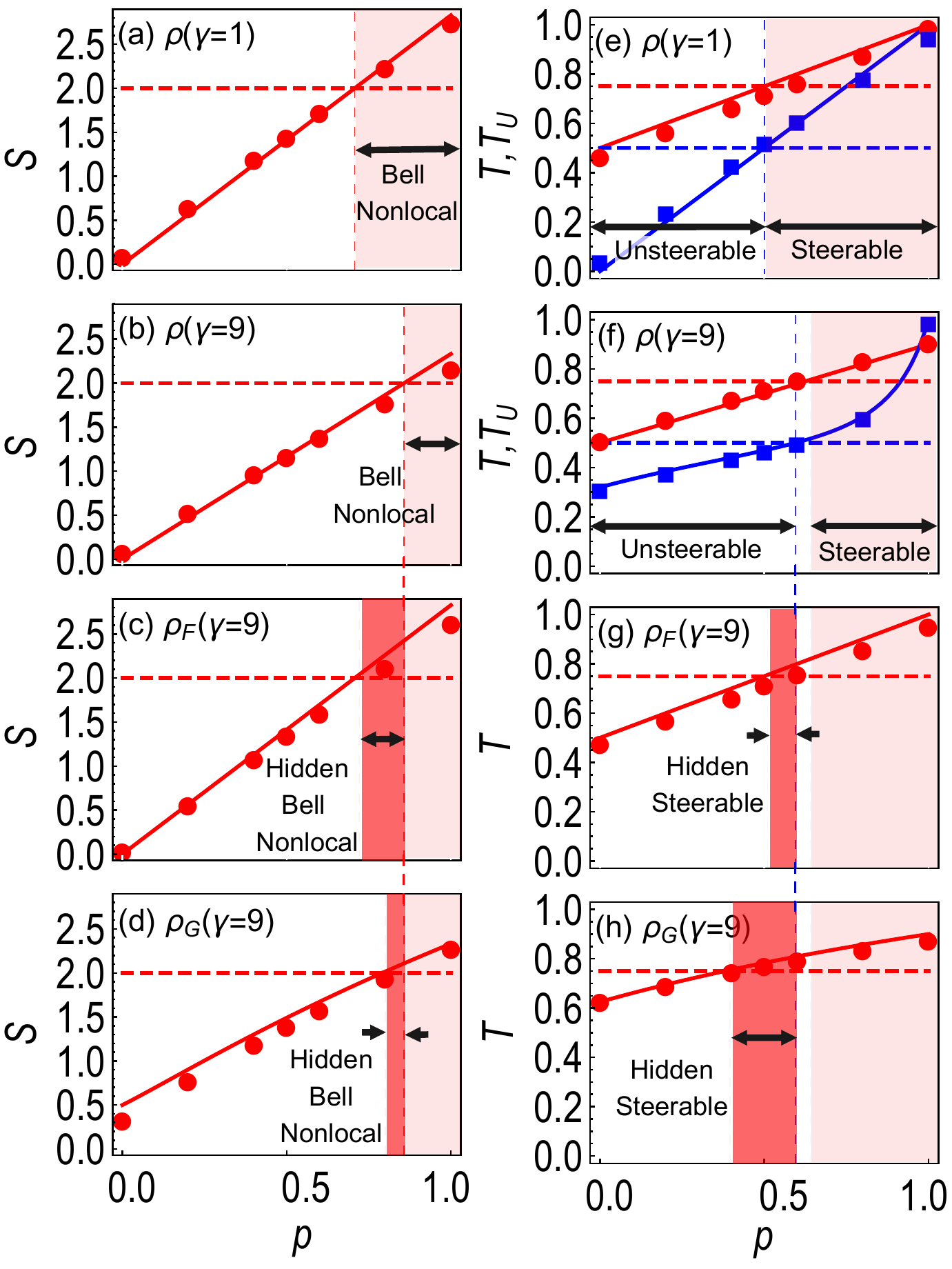}
\caption{ Theoretical and experimental results  of (a)-(d) Bell nonlocality and (e)-(h) steerability. Here, red and blue marks (lines) correspond to experimental (theoretical) data (prediction) of respective correlations {\color{black}for $p=0, \,0.2, \,0.4, \,0.5, \,0.6, \,0.8, \,1$.} The light shaded regions represent Bell-nonlocal regions for (a) $\gamma=1$, and (b)-(d) $\gamma=9$, respectively. The revealed hidden Bell-nonlocal ranges after the local filtering operations are presented as dark shaded regions in (c) and (d). The light shaded regions represent steerable regions for (e) $\gamma=1$, and (f)-(h) $\gamma=9$, respectively. The revealed hidden steerable ranges after the local filtering operations are presented as dark shaded regions in (c) and (d). Error bars are smaller than the size of markers.}
\label{data_Steer}
\end{figure}

We present the theoretical and experimental Bell parameter $S$ of the investigated states with respect to the parameter $p$ in Figs.~\ref{data_Steer}\,(a)-(d). The horizontal straight lines correspond to the upper bound of $S=2$ of Bell inequality under LHV model.

Bell nonlocality of $\rho$ for $\gamma=1$ is shown in the Fig.~\ref{data_Steer}\,(a). The light shaded area displays Bell nonlocal region with respect to the parameter $p$. It clearly presents that the state is Bell nonlocal for $p>1/\sqrt{2}$ both in theory and experiment. 

Figures~\ref{data_Steer}\,(b)-(d) show the Bell parameter $S$ of the state $\rho$, $\rho_F$, and $\rho_G$ for $\gamma=9$, respectively. Here, the light shaded regions correspond to the Bell nonlocal region of $\rho$ for $\gamma=9$. Figures~\ref{data_Steer}\,(c) and (d) show that both local filters $\mathcal{F}_{A,B}$ and $\mathcal{G}_{A,B}$ reveal hidden Bell nonlocality, see the dark shaded regions. As presented by the area of the dark shaded regions, $\mathcal{F}_{A,\,B}$ are more effective than $\mathcal{G}_{A,\,B}$ to uncover Bell nonlocality. In the experiment, $\mathcal{G}_{A,\,B}$ fail to uncover hidden Bell nonlocality for $p=0.8$ while $\mathcal{F}_{A,\,B}$ successfully find it. It is noteworthy to remark that the region of hidden Bell nonlocality can be as large as that of the state $\rho$ with $\gamma=1$ (i.e., mixed states originating from maximally entangled state) with the local filters $\mathcal{F}_{A,B}$, and never exceed this limit, see Fig.~\ref{data_Steer}\,(a) and (c). It means that the hidden Bell nonlocality boundaries of mixed states from non-maximally pure entangled states can be stretched to the Bell nonlocality boundary of the mixed state from maximally entangled states. 

The theoretical and experimental results of steerability and unsteerability are presented in Figs.~\ref{data_Steer}\,(e) and (f). Experimental data and theoretical prediction correspond to red marks and lines to denote steering parameter $T$ of Eq.~(\ref{FUR_St}). Whereas blue marks and lines present experimental data obtained by constructing the state with the help of quantum~state~tomography  and theoretical prediction of unsteering parameter $T_U$ of Eq.~(\ref{Steer_Bowel}). The horizontal red line corresponds to the upper bound of the steering inequality, and thus, the state is steerable above this line. On the other hand, the horizontal blue line denotes the upper bound of the sufficient criterion for unsteerability, i.e., $T_U=1/2$. Therefore, according to the criterion of Eq.~(\ref{Steer_Bowel}), the state is unsteerable below this line. In order to present unsteerable region with respect to the parameter $p$, we present vertical blue lines at the intersection of theoretical $T_U$ curves and $T_U=1/2$.

The light shaded regions in Figs.~\ref{data_Steer}\,(e) and (f) present the range of $p$ for which the state $\rho$ with {\color{black}$\gamma=1$ and $\gamma=9$} are steerable, respectively. The Fig.~\ref{data_Steer}\,(e) shows that the state with $\gamma=1$ is unsteerable for $p\leq1/2$ and steerable for $p>1/2$. Therefore, the steering criterion Eq.~(\ref{FUR_St}) becomes necessary and sufficient for $\rho$ with $\gamma=1$. In the case of $\gamma=9$, there exists a gap of $0.595<p\leq0.625$ between steerable and unsteerable regions. Note that the steerability or unsteerability of $\rho$ cannot be concluded in this region.


The results of steerability of $\rho_{\mathcal{F}}$ and $\rho_{\mathcal{G}}$ for $\gamma=9$ are presented in Figs.~\ref{data_Steer}\,(g) and (h), respectively.  To visualize hidden steerable region, we present the same vertical blue line of Fig.~\ref{data_Steer}\,(f) which guarantees unsteerability of $\rho$ for $\gamma=9$. The hidden steerable regions that are shaded by dark red can be clearly seen both in Figs.~\ref{data_Steer}\,(g) and (h). Since the hidden steerable region of the Fig.~\ref{data_Steer}\,(h) is larger than that of Fig.~\ref{data_Steer}\,(g), $\mathcal{G}_{A,\,B}$ are more effective than $\mathcal{F}_{A,\,B}$ for uncovering hidden steerability. In experiment, $\mathcal{G}_{A,\,B}$ successfully uncover hidden steerability for $p=0.5$, however $\mathcal{F}_{A,\,B}$ fail. In addition, we note that there exists wide region which shows hidden steerability, but fails to show hidden Bell nonlocality, see Figs.~\ref{data_Steer}\,(d) and (h). 


It is remarkable that the steerability boundaries of $\rho_{\mathcal{G}}$ with $\gamma=9$ can reach smaller $p$, meaning more noise, than that of $\rho$ with $\gamma=1$, see Figs.~\ref{data_Steer}\,(e) and (h). In particular, for $p=0.5$, $\rho_{\mathcal{G}}$ with $\gamma=9$ state successfully reveals steerability whereas $\rho$ with $\gamma=1$ fails. This experimental result supports our theoretical finding that mixed states from non-maximally pure entangled states can have hidden steerability for wider range of the noise parameter than the mixed state originating from a maximally pure entangled state.

\section{Conclusion}

In conclusion, we have theoretically and experimentally investigated {\color{black} revealing of steerability of entangled states having LHS model} with the help of local filtering operations. As LOCC can not create nonlocal correlations, steerability was hidden in the initial states. We have proven that hidden steerability is distinguishable from hidden Bell nonlocality by showing that there are quantum states which reveal hidden steerability, but fail to show hidden Bell nonlocality. We have investigated two sets of local filters and found that one set of filters is more effective on revealing hidden Bell nonlocality whereas the other is more effective on uncovering hidden steerability. {\color{black} Therefore, EPR steerability shows different characteristic from  Bell nonlocality  under unitary operations. Note that Bell nonlocality impels steerability.}

We have also presented a counter-intuitive result that mixed states from non-maximally pure entangled states can have hidden steerability even when the mixed state originating from a maximally entangled state does not have steerability. Considering the fundamental importance and applications of quantum steerability in quantum information science, our findings provide a better understanding of nonlocal quantum correlation and paves the way towards secure quantum communications.

\section*{Acknowledgements}





\end{document}


\onecolumngrid

{\Large\bf Supplementary materials}

\section{Calculation of entanglement}

To measure the entanglement of a bipartite state, we calculate concurrence of the sate. If concurrence is positive, then the state is entangled. The concurrence of the state $\rho$ can be calculated from the eigenvalues of $\Lambda^C=\rho\cdot\left(\sigma_y\otimes\sigma_y\cdot\rho^*\cdot\sigma_y\otimes\sigma_y\right)$ where the asterisk `$*$' stands for complex conjugation. For the state $\rho$, the eigenvalues of $\Lambda^C$ in decreasing order becomes
\begin{eqnarray}
\lambda_1 &=& \frac{1}{16} (3 p+1)^2 \sin^2 2 \theta, \nonumber \\
\lambda_2 &=& \lambda_3=\lambda_4=\frac{1}{16} (1-p)^2 \sin^2 2 \theta .
\end{eqnarray}
The concurrence of the state $\rho$ is
\begin{eqnarray}
C^{\rho} &=& \max\left[\sqrt{\lambda_1} - \sqrt{\lambda_2} - \sqrt{\lambda_3} - \sqrt{\lambda_4},0\right]\nonumber \\
 &=& \max\left[(3 p-1) \sin \theta \cos \theta,0\right].
\end{eqnarray}
For the Werner states $\rho_{\mathcal{F}}$, $C^{\rho_{\mathcal{F}}} = \frac{3p-1}{2}$ since $\theta=\frac{\pi}{4}$.

Similarly, one can calculate concurrence of the sate $\rho_{\mathcal{G}}$. The eigenvalues of $\Lambda^C$ in decreasing order are
\begin{eqnarray}
\lambda_1 &=& \left[\frac{3 p+1}{(p+1) \csc \theta  \sec\theta -2 p+2} \right]^2, \nonumber \\
\lambda_2 &=& \lambda_3=\lambda_4=\frac{1-p}{(p+1) \csc \theta  \sec\theta -2 p+2}.
\end{eqnarray} 
Therefore, concurrence of the state $\rho_{\mathcal{G}}$ is
\begin{eqnarray}
C^{\rho_{\mathcal{G}}} = \max\left[\frac{6 p-2}{(p+1) \csc \theta  \sec\theta -2 p+2},0\right].
\end{eqnarray}

{\color{black}
\section{Calculation of Bell nonlocality and corresponding measurement settings}

To obtain maximal Bell violation of a given state $\sigma$, we have used Horodecki criterion~\cite{Horo_Cri, Horo_Cri_2}. Note that, for the scenario of two measurement settings on each $2\times 2$ and $2\times 3$ dimensional systems, the Horodecki criterion is necessary and sufficient for testing Bell nonlocality. According to the Horodecki criterion, the maximum Bell violation of $\sigma$ can be calculated form the correlation matrix $\Lambda^\sigma$ of the form
\begin{eqnarray}
\Lambda^\sigma_{ij} = Tr[\sigma_i\otimes \sigma_j\cdot\sigma],
\end{eqnarray} 
and it becomes
\begin{eqnarray}
S^\sigma=2\sqrt{\Gamma_1^\sigma +\Gamma_2^\sigma},
\end{eqnarray} 
where $\Gamma_1^\sigma$ and $\Gamma_2^\sigma$ are two maximum eigenvalues of $\Lambda^\sigma$.  Note that the state $\sigma$ is called Bell nonlocal when $S^\sigma>2$.

For the considered state $\rho$, the eigenvalues of $\Lambda^\rho$ in decreasing order are given by
\begin{eqnarray}
\Gamma_1^\rho=p^2\\
\Gamma_2^\rho=p^2 \, \sin^22\theta,
\end{eqnarray}
and the maximum Bell violation of the state $\rho$ becomes
\begin{eqnarray}
S^\rho=2\,p\,\sqrt{1+\sin^22\theta}.
\label{S_rho}
\end{eqnarray}
Therefore, $\rho$ is Bell nonlocal for $p\,\sqrt{1+\sin^22\theta}>1$. 
In the case of Wener state $\rho_{\mathcal{F}}=\frac{(\mathcal{F}_A\otimes \mathcal{F}_B)\cdot\rho\cdot(\mathcal{F}_A^\dagger\otimes \mathcal{F}_B^\dagger)}{\Tr[(\mathcal{F}_A\otimes \mathcal{F}_B)\cdot\rho\cdot(\mathcal{F}_A^\dagger\otimes \mathcal{F}_B^\dagger)]}=\rho_{\theta=\pi/4}$ $S^\rho_{\theta=\pi/4}=2\,\sqrt{2}\,p$ and it is Bell nonlocal for $p>\frac{1}{\sqrt{2}}$.

To obtain the measurement settings for Alice and Bob corresponding to the Bell violation given by Eq.~(\ref{S_rho}) for the shared state $\rho$, Alice measures either observables $\mathcal{A}_1=\sigma_x$ or $\mathcal{A}_2=\sigma_z$ on her system $A$. Bob's choice of observables are
\begin{eqnarray}
\mathcal{B}_1 &=&\sigma_z\cos\varphi_1 +\sigma_x\sin\varphi_1, \nonumber \\
\mathcal{B}_2&=&\sigma_z\cos\varphi_2 +\sigma_x\sin\varphi_2.
\label{BN_MS_rho}
\end{eqnarray}
The value of Bell parameter $S$ becomes
\begin{eqnarray}
S^{\rho} &=& \Tr\left[\left(\mathcal{A}_1\,\mathcal{B}_1 +\mathcal{A}_1\,\mathcal{B}_2 + \mathcal{A}_2\,\mathcal{B}_1 -\mathcal{A}_2\,\mathcal{B}_2 \right) \cdot \rho \right]\nonumber \\
&=& p \left[\sin 2 \theta  (\sin \varphi_1+\sin \varphi_2)+\cos \varphi_1-\cos \varphi_2\right]. 
\end{eqnarray}

The maximum value of $S^{\rho}$ given by Eq.~(\ref{S_rho}) can be found when $\cos\varphi_1=-\cos\varphi_2=\frac{1}{\sqrt{1+\sin^22\theta}}$, $\sin\varphi_1>0$, and $\sin\varphi_2 > 0$. Therefore, Bob's measurement settings corresponding to the maximum value of Bell parameter $S^{\rho}$ are
\begin{eqnarray}
\varphi_1 &=& \pi + \arctan\left[ \sin^2 2\theta\right]=\pi+\arctan\left[\sqrt{4\gamma/(1+\gamma)^2}\right]
\nonumber \\
\varphi_2 &=& \arctan\left[- \sin^2 2\theta\right]=\arctan\left[-\sqrt{4\gamma/(1+\gamma)^2}\right].
\label{theta12}
\end{eqnarray}
In the case of Werner state $\rho_{\mathcal{F}}$, or equivalently $\theta=\frac{\pi}{4}$, Bob's choice of measurement settings become $\mathcal{B}_1=\frac{\sigma_x+\sigma_z}{\sqrt{2}}$ and $\mathcal{B}_2=\frac{\sigma_x-\sigma_z}{\sqrt{2}}$, respectively.

Similarly, the for the state $\rho_{\mathcal{G}}=\frac{(\mathcal{G}_A\otimes \mathcal{G}_B)\cdot\rho\cdot(\mathcal{G}_A^\dagger\otimes \mathcal{G}_B^\dagger)}{\Tr[(\mathcal{G}_A\otimes \mathcal{G}_B)\cdot\rho\cdot(\mathcal{G}_A^\dagger\otimes \mathcal{G}_B^\dagger)]}$, the eigenvalues of $\Lambda^\rho_{\mathcal{G}}$ in decreasing order are given by
\begin{eqnarray}
\Gamma_1^{\rho_{\mathcal{G}}}&=&\frac{16 p^2}{(2 - 2p + \csc\theta\,\sec\theta + p \csc\theta\,\sec\theta)^2}\\
\Gamma_2^{\rho_{\mathcal{G}}}&=&\left(\frac{-2 + 2p + \csc\theta\,\sec\theta + p \csc\theta\,\sec\theta}{2 - 2p + \csc\theta\,\sec\theta + p \csc\theta\,\sec\theta}\right)^2.
\end{eqnarray}
The maximum Bell violation of the state $\rho^{\mathcal{G}}$ becomes
\begin{eqnarray}
S^\sigma=2\sqrt{\Gamma_1^{\rho_{\mathcal{G}}} +\Gamma_2^{\rho_{\mathcal{G}}}}.
\label{S_rho_G}
\end{eqnarray}
Therefore, the state $\rho_{\mathcal{G}}$  is Bell nonlocal for  $\left(p^2-1\right) \csc \theta  \sec \theta +5 p^2+(p+1)^2 \csc ^2 2 \theta -2 p+1>\frac{1}{2}[(p+1) \csc \theta \sec \theta -2 p+2]^2$. 

The value of Bell parameter $S$ for the state $\rho_{\mathcal{G}}$ with the set of Alice's Bob's observables $\{\mathcal{A}_1=\sigma_x,\, \mathcal{A}_2=\sigma_z\}$, and $\mathcal{B}_{1,2}$ given by Eq.~(\ref{BN_MS_rho}) becomes
\begin{eqnarray}
S^{\rho_{\mathcal{G}}} = \frac{8 \left(\left(p^2-1\right) \csc\theta \sec\theta +(p+1)^2 \csc^22\theta+p (5 p-2)+1\right)}{\left(4 (p+1)^2 \csc ^22 \theta -4 (p-1)^2\right) \sqrt{\frac{16 p^2}{((p+1) \csc\theta \sec\theta +2 (p-1))^2}+1}}.
\end{eqnarray}

It is not difficult to obtain the Bob's observables corresponding to maximum value of the Bell parameter $S^{\rho_{\mathcal{G}}}$ given by Eq.~(\ref{S_rho_G}) as
\begin{eqnarray}
\varphi_1 &=& \pi + \arctan\left( \frac{4 p}{(p+1) \csc\theta  \sec\theta +2 (p-1)}\right) = \pi + \arctan\left[ \frac{4 p}{\sqrt{\frac{1}{\gamma }} (\gamma +1)(p+1)+2 (p-1)}\right]
,\nonumber\\
\varphi_2 &=& \arctan\left(- \frac{4 p}{(p+1) \csc\theta  \sec\theta +2 (p-1)}\right)=\arctan\left[ -\frac{4 p}{\sqrt{\frac{1}{\gamma }} (\gamma +1)(p+1)+2 (p-1)}\right].
\end{eqnarray}

}

\section{Calculation of steerability}

For the observables of $\mathcal{A,B}\in\{\sigma_z,\,\sigma_x\}$, the conditional probability distribution $P(b=0_{\mathcal{B}}|a=0_{\mathcal{A}})$ for the shared state $\rho$ becomes
\begin{eqnarray}
P(0_{\sigma_z}|0_{\sigma_z}) &=& \frac{1+p}{2}, \nonumber \\
P(0_{\sigma_x}|0_{\sigma_x}) &=& \frac{1+p \sin2\theta}{2}.
\end{eqnarray}
Therefore, the steering parameter $T$ given by Eq.(17) for the outcomes $a=b=0$ becomes
\begin{eqnarray}
T^{\rho}&=&\frac{1}{2}\sum_{i=z,x}P(b=0_{\sigma_i}|a=0_{\sigma_i}) \nonumber \\
&=& \frac{2+p+p\sin2\theta}{4}.
\label{T_rho}
\end{eqnarray}
The value of the steering parameter $T$ for Werner state $\rho_{\mathcal{F}}$ can be obtained from Eq.~(\ref{T_rho}) for the choice $\theta=\pi/4$, and it becomes
\begin{eqnarray}
T^{\rho_{\mathcal{F}}}=\frac{1+p}{2}.
\end{eqnarray}

Similarly, the conditional probability  distribution $P(b=0_{\mathcal{B}}|a=0_{\mathcal{A}})$  for the shared state $\rho_{\mathcal{G}}$ becomes
\begin{eqnarray}
P(0_{\sigma_z}|0_{\sigma_z}) &=& \frac{(p+1) \cot \theta }{(p+1) \cot\theta -p+1}, \nonumber \\
P(0_{\sigma_x}|0_{\sigma_x}) &=& \frac{2 p}{(p+1) \csc \theta  \sec\theta -2 p+2}.
\end{eqnarray}
Therefore, the steering parameter $T$ of the state $\rho_{\mathcal{G}}$ for the outcomes $a=b=0$ becomes
\begin{eqnarray}
T^{\rho_{\mathcal{G}}}&=&\frac{1}{2}\sum_{i=z,x}P(b=0_{\sigma_i}|a=0_{\sigma_i})\nonumber \\
&=& \frac{1}{2-\frac{2 (p-1) \tan \theta}{p+1}} +\frac{p}{(p+1) \csc\theta  \sec \theta -2 p+2}.
\label{T_rho_G}
\end{eqnarray}
\section{Experimental setup and data analysis}
\begin{figure}[b]
\includegraphics[width=4 in]{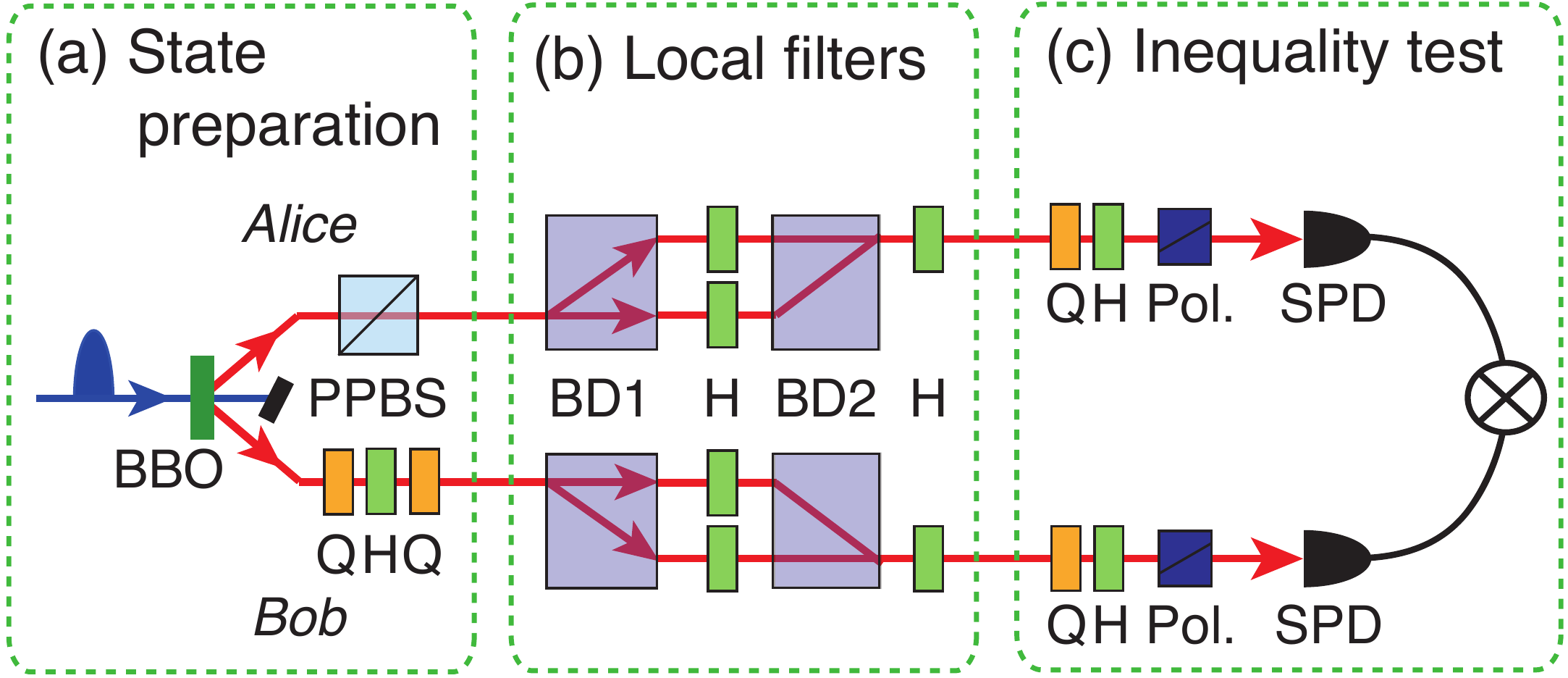}
\caption{Experimental setup. PPBS: partially polarizing beamsplitter, Q: quarter waveplate, H: half waveplate, BD: beam displacer, Pol.: polarizer, SPD: single-photon detector. See text for details.
}
\label{setup}
\end{figure}

Figure~\ref{setup} shows the experimental setup to explore different hidden quantum correlations. First, let us describe the initial state preparation as shown in Fig.~\ref{setup} (a). Maximally polarization entangled photon pairs of $|\psi\rangle=\frac{1}{\sqrt{2}}(|00\rangle+|11\rangle)=\frac{1}{\sqrt{2}}(|HH\rangle+|VV\rangle)$ centered at 780~nm are generated at a sandwich BBO crystal via the SPDC process pumped by a femtosecond pump pulse centered at 390~nm. Here, $|H\rangle$ and $|V\rangle$ denote horizontal and vertical polarization states, respectively. The sandwich BBO crystal, which is composed of two type-II BBO crystals and a half waveplate, is specially designed for efficient generation of two-photon entangled states~\cite{wang16}. Each of the entangled photon pairs is then sent to Alice and Bob, respectively.

The parameter $\theta$ or equivalently $\gamma$ is controlled by partially polarizing beamsplitters (PPBS) at Alice. The transmitivity of horizontal and vertical polarization states of the PPBS are $T_H=1$ and $T_V=\frac{1}{3}$, respectively. The depolarizing channel can be implemented by statistically mixing identity $(I)$ and three Pauli operations $(\sigma_x,\sigma_y,\sigma_z)$ with a set of waveplates (QHQ) placed at Bob~\cite{lim11,lim11b,lim12}. The noise parameter $p$ is determined by adjusting the ratio of the single-qubit operations during the data acquisition.

Since we use the single-photon polarization qubits, the local filters $\mathcal{F}_{1,2}$ and $\mathcal{G}_{1,2}$ correspond to polarization dependent loss~\cite{kim09}. As depicted in Fig.~\ref{setup}(b), the local filtering operations are implemented by a set of beam displacers (BD) and half waveplates (HWP) for high phase stability~\cite{kim11}. While the vertically polarized beam is transmitted through the BD1, the horizontal polarization state is displaced after the BD1. Then, the polarization state at each arm is flipped by the following HWP at $45\degree$, and two beams are combined together at BD2. The amount of the polarization dependent loss can be adjusted by changing the angles of HWPs between two BDs. In order to restore the flipped polarization state, we put a HWP at $45\degree$ after the second BD.

The experimental tests of Bell's and steering inequalities require various two-qubit projection measurements, which can be implemented by waveplates, polarizers, and single-photon detectors as shown in Fig.~\ref{setup}(c). The coincidence detection between Alice and Bob is achieved by a home-made coincidence counting unit~\cite{park15}. 

{\color{black}
In the experiment, the unsteering parameter $T_U$ is obtained from state tomography and the parameters $S$ and $T$ are calculated from the  joint probability distribution $P(a_{\mathcal{A}}, b_{\mathcal{B}})$. The joint probability distribution $P(a_{\mathcal{A}}, b_{\mathcal{B}})$ are experimentally estimated from coincidence detection of photons at two single photon detectors as
\begin{eqnarray}
P(a_{\mathcal{A}}, b_{\mathcal{B}}) &=& \frac{C(a_\mathcal{A},b_{\mathcal{B}}) }{C(0_\mathcal{A},0_{\mathcal{B}}) +  C(0_\mathcal{A},1_{\mathcal{B}}) + C(1_\mathcal{A},0_{\mathcal{B}}) + C(1_\mathcal{A},1_{\mathcal{B}}) }, \\
P(a_{\mathcal{A}}) &=& \frac{C(a_\mathcal{A},0_{\mathcal{B}}) + C(a_\mathcal{A},1_{\mathcal{B}})}{C(0_\mathcal{A},0_{\mathcal{B}}) +  C(0_\mathcal{A},1_{\mathcal{B}}) + C(1_\mathcal{A},0_{\mathcal{B}}) + C(1_\mathcal{A},1_{\mathcal{B}}) },
\end{eqnarray}
where $C(a_{\mathcal{A}}, b_{\mathcal{B}})$ represents coincidence counts for outcome $a\in\{0,\,1\}$ and $b\in\{0,\,1\}$. 

Therefore, steering parameter $T$ in terms of coincidence detection becomes
\begin{eqnarray}
T & =&\frac{1}{2}\left(P(b_{\sigma_x}|a_{\sigma_x}) + P(b_{\sigma_z}|a_{\sigma_z})\right)= \frac{1}{2}\left(\frac{P(a_{\sigma_x}, b_{\sigma_x})}{P(a_{\sigma_x})} + \frac{P(a_{\sigma_z}, b_{\sigma_z})}{P(a_{\sigma_z})}\right) \nonumber \\
& = & \frac{C(a_{\sigma_x},b_{\sigma_x})}{C(a_{\sigma_x},0_{\sigma_x}) + C(a_{\sigma_x},1_{\sigma_x})} + \frac{C(a_{\sigma_z},b_{\sigma_z})}{C(a_{\sigma_z},0_{\sigma_z}) + C(a_{\sigma_z},1_{\sigma_z})}
\end{eqnarray}
The expectation values of $\langle \mathcal{A}_i\,\mathcal{B}_j \rangle$ where $i,j\,\in\{0,\,1\}$ are calculated as
\begin{eqnarray}
\langle \mathcal{A}_i\,\mathcal{B}_j \rangle =  \frac{ C(0_{\mathcal{A}_i},0_{\mathcal{B}_j}) -  C(0_{\mathcal{A}_i},1_{\mathcal{B}_j}) - C(1_{\mathcal{A}_i},0_{\mathcal{B}_j}) + C(1_{\mathcal{A}_i},1_{\mathcal{B}_j}) }{ C(0_{\mathcal{A}_i},0_{\mathcal{B}_j}) +  C(0_{\mathcal{A}_i},1_{\mathcal{B}_j}) + C(1_{\mathcal{A}_i},0_{\mathcal{B}_j}) + C(1_{\mathcal{A}_i},1_{\mathcal{B}_j}) } = \frac{\displaystyle\sum_{a,b=0}^{1} (-1)^{(a+b)} \,C(a_{\mathcal{A}_i}, b_{\mathcal{B}_j}) }{\displaystyle\sum_{a,b=0}^{1}  \,C(a_{\mathcal{A}_i}, b_{\mathcal{B}_j}) }.
\end{eqnarray}
The Bell parameter in terms of coincidence counts becomes
\begin{eqnarray}
S &=&\langle \mathcal{A}_1\,\mathcal{B}_1 \rangle + \langle \mathcal{A}_1\,\mathcal{B}_2 \rangle +\langle \mathcal{A}_2\,\mathcal{B}_1 \rangle -\langle \mathcal{A}_2\,\mathcal{B}_2 \rangle \nonumber \\
&=& \frac{\displaystyle\sum_{a,b=0}^{1} (-1)^{(a+b)} \,C(a_{\mathcal{A}_1}, b_{\mathcal{B}_1}) }{\displaystyle\sum_{a,b=0}^{1}  \,C(a_{\mathcal{A}_1}, b_{\mathcal{B}_1}) } 
+ \frac{\displaystyle\sum_{a,b=0}^{1} (-1)^{(a+b)} \,C(a_{\mathcal{A}_2}, b_{\mathcal{B}_2}) }{\displaystyle\sum_{a,b=0}^{1}  \,C(a_{\mathcal{A}_2}, b_{\mathcal{B}_2}) } 
+ \frac{\displaystyle\sum_{a,b=0}^{1} (-1)^{(a+b)} \,C(a_{\mathcal{A}_3}, b_{\mathcal{B}_3}) }{\displaystyle\sum_{a,b=0}^{1}  \,C(a_{\mathcal{A}_3}, b_{\mathcal{B}_3}) } \nonumber \\
&& - \frac{\displaystyle\sum_{a,b=0}^{1} (-1)^{(a+b)} \,C(a_{\mathcal{A}_4}, b_{\mathcal{B}_4}) }{\displaystyle\sum_{a,b=0}^{1}  \,C(a_{\mathcal{A}_4}, b_{\mathcal{B}_4}) }
\end{eqnarray}

In our experiment, we have prepared the initial Bell state with the fidelity of $97\%$. The typical coincidence counts are about $2\times10^3$ with the accidental coincidence counts of $\sim10$ for 5 second of data acquisition. We have taken 10 sets of data and have calculated experimental standard deviation as a measure of error.  $10^{-2}$  and $10^{-3}$  are the order of error bars for the Bell and steering parameters, respectively.
}